\documentstyle[11pt,newpasp,twoside,epsf]{article}
\def\edcomment#1{\iffalse\marginpar{\raggedright\sl#1\/}\else\relax\fi}
\reversemarginpar

\begin{document}
\title{Galaxy structure and kinematics towards the NGP}
 \author{Spagna Alessandro}
\affil{INAF-Osservatorio Astronomico di Torino, I-10025 Pino
Torinese, Italy}
\author{Cacciari Carla}
\affil{INAF - Osservatorio Astronomico di Bologna, I-40127
Bologna}
\author{Drimmel Ronald}
\affil{INAF-Osservatorio Astronomico di Torino, I-10025 Pino
Torinese, Italy}
\author{Kinman Thomas}
\affil{Kitt Peak National Observatory, NOAO, Tucson, AZ
85726-6732, USA }
\author{Lattanzi Mario G.}
\affil{INAF-Osservatorio Astronomico di Torino, I-10025  Pino
Torinese, Italy}
\author{Smart Richard L.}
\affil{INAF-Osservatorio Astronomico di Torino, I-10025  Pino
Torinese, Italy}

\begin{abstract}
We present a proper motion survey over about 200 square degrees
towards the NGP, based on the material used for the construction
of the GSC-II, that we are using to study the vertical structure
and kinematics of the Galaxy.  In particular, we measured the
rotation velocity of the halo up to 10 kpc above the galactic
plane traced by a sample of RR Lyr{\ae} and BHB giants for which
radial velocities were used to recover the complete distribution
of the spatial velocities.  Finally, the impact of astrometric and
spectroscopic GAIA observation are discussed.
\end{abstract}

\section{Introduction}
It is generally accepted that the Galaxy is constituted by four
discrete main components, the {\it bulge}, the {\it thin disk},
the {\it thick disk} and the {\it halo}, which are characterized
by distinctive stellar populations in terms of spatial
distribution, kinematics properties, metallicity and age. A
detailed knowledge of such galactic components is essential to
achieve a complete description of the Milky Way, as well as of the
various processes and evolutionary phases which occurred during
the history of our Galaxy (and other galaxies too) and that are
responsible for the existence and the properties of those
components we observe today.

Ground based surveys providing photometry, proper motions and/or
spectroscopic observations have been carried out in the past to
study the structure and kinematics of Galactic populations, and
these will continue and be extended in the next years thanks to
availability of all-sky catalogs such as GSC-II and USNO-B, based
on multi-epoch photographic surveys, as well as other current and
future photometric and spectroscopic surveys (eg.\ 2MASS, DENIS,
SDSS, EIS, HES, HK, VST, etc.) that  benefit from the availability
of dedicated scanning and large field cameras 
as well as  large multi-fiber spectrographs (eg. 2dF, 6dF,
FLAMES).

Clearly, the GAIA mission will provide an enormous contribution to
the understanding of the formation and evolution of the Galaxy,
thanks to very accurate astrometric parameters complemented by
photometric and spectroscopic measurements which will permit the
direct determination of the 6D phase space distribution and the
chemical abundance of large and complete samples of stellar
tracers belonging to the different galactic components.

Here we describe a new project which combines astrometric,
photometric and spectroscopic data in order to investigate the
kinematics of the outer halo by means of velocities derived from
proper motions and radial velocities for a sample of RR-Lyr{\ae}
and BHB giants towards the NGP.

\section{NGP survey: proper motions and radial velocities}
At the moment we have surveyed about 200 square degrees towards
the NGP, and produced positions, proper motions and photographic
photometry for about 500\,000 objects down to plate limits (R$_F <
20.5$).

\begin{table}
  \centering
  \caption{Plate material}
  \smallskip
\begin{tabular}{rccll}
\tableline
            &          &     &     &   \\
{\it Survey} & {\it Epoch} & {\it Pixel} & {\it Band}  &{\it
Emulsion + Filter} \\
            &          &     &     &   \\
  \tableline
            &          &     &     &   \\
 POSS-I E   &  1950-1956 & 25 $\mu$m & E   & 103a-E + red plexiglass\\
 POSS-I O   &  1950-1955 & 25 $\mu$m & O   & 103a-O unfiltered \\
 Quick V   &  1982-1983 &  25 $\mu$m & V$_{12}$ & IIaD+Wratten 12 \\
 POSS-II J &  1988-1996 &  15 $\mu$m & B$_J$ & IIIaJ + GG385 \\
 POSS-II F  &  1989-1996 & 15 $\mu$m & R$_F$ & IIIaF + RG610 \\
 POSS-II N  &  1990-1998 & 15 $\mu$m & I$_N$ & IV-N + RG9    \\
            &          &     &    & \\
 \tableline
\end{tabular}

  \label{}
\end{table}

  Our material consists of  $6.4^\circ\times 6.4^\circ$
Schmidt plates from the Northern photographic surveys (POSS-I,
Quick V and POSS-II) carried out at the Palomar Observatory (Table
1).  All plates were digitized at STScI utilizing modified
PDS-type scanning machines with 25 $\mu$m square pixels (1.7
$''$/pixel) for the first epoch plates, and 15 $\mu$m pixels (1
$''$/pixel) for the second epoch plates. The digital copies of the
plates were initially analyzed by means of the standard software
pipeline used for the construction of the GSC-II (see e.g.\ Lasker
et al.\ 1995 or McLean et al.\ 2000). The pipeline performs object
detection and computes parameters and features for each identified
object. Further, the software provides classification, position,
and magnitude for each object by means of astrometric and
photometric calibrations which utilized the Tycho2 (H{\o}g et al.
2000) and the GSPC-2 (Bucciarelli et al. 2001) as reference
catalogs. Accuracies better than 0.1-0.2 arcsec in position and
0.15-0.2 mag in photographic magnitude are generally attained.
Relative proper motions were derived by applying the procedure
described in Spagna et al. (1996) and afterwards transformed to
the absolute reference frame forcing the extended extragalactic
sources to have null tangential motion. As shown in Figure 1, the
typical precision ($\sigma_\mu\sim 3$ mas/yr down to $R_F\simeq
18$) has been estimated from the formal errors of the fitted
proper motions, while the zero point accuracy of the absolute
proper motions have been tested by checking the mean motion of a
set of known QSO's that give values smaller than 1 mas/yr on each
component.

\begin{figure}
 \plotfiddle{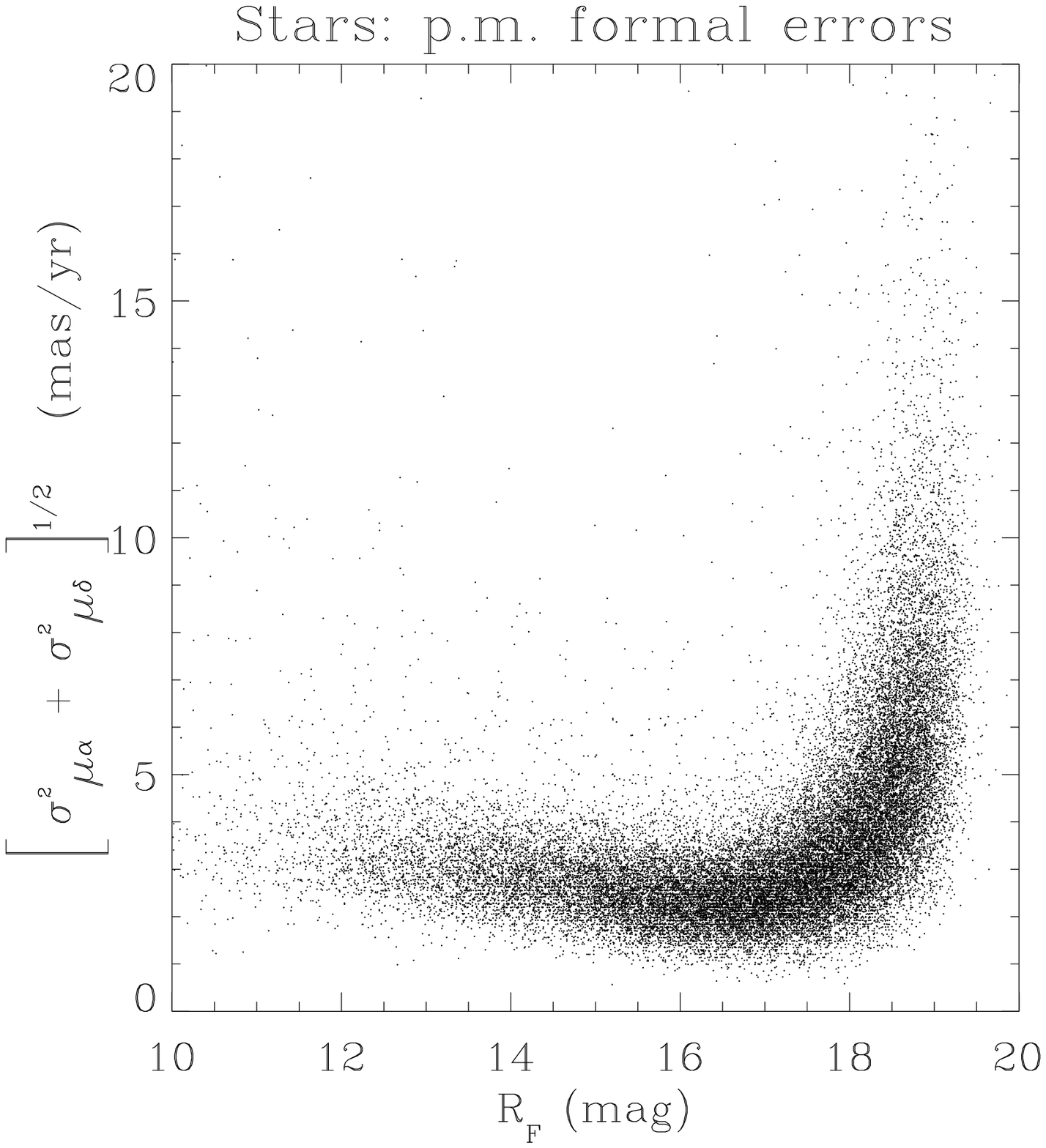}{7cm}{0}{45}{45}{-220}{-20}
 \plotfiddle{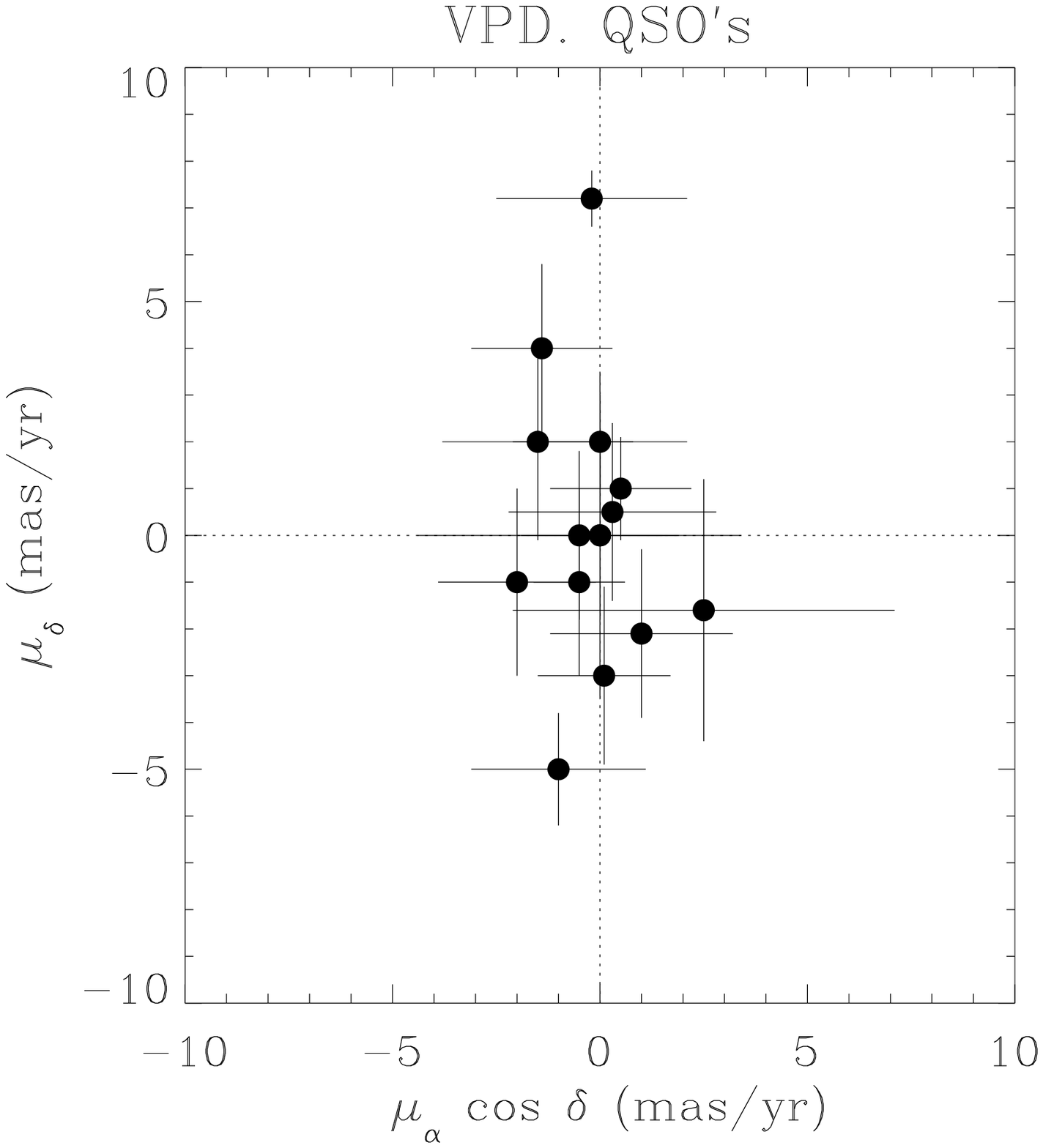}{0cm}{0}{45}{45}{-10}{0}
 \caption{{\it Left panel}. Formal errors of fitted proper
 motions as a function of the magnitude (all stellar objects).
 {\it Right panel.} Vector point diagram (VPD) of a sample of 15
 QSO's, with 1 $\sigma$ error bars.  Weighted means are $\langle \mu_\alpha \cos\delta\rangle = -0.02 \pm 0.23$
 mas/yr and $\langle\mu_\delta\rangle = +0.33 \pm 0.28$ mas/yr, respectively.
 Both plots refer to the POSS-II field no.\ 442.}
\end{figure}

 Radial velocities and chemical abundances of the sample
of RR Lyr{\ae} and BHB giants were derived by means of
spectroscopic observations carried out with the 4m Mayall
telescope at Kitt Peak and with the 3.5m TNG on La Palma.  The
data were processed with standard procedures and routines (IRAF),
and typical errors are $\sigma_{\rm RV} \le 40$ km s$^{-1}$ and
$\sigma_{\rm [Fe/H]}\sim 0.2$ dex.

\section{The vertical structure}
The color magnitude diagram and the vector point diagram observed
in one field of our NGP survey are shown in Figure 2.  The
observed distributions are the result of the complex mixture of
the stars belonging to the various populations which are present
towards high galactic latitudes. In particular, these may include:
\begin{enumerate}
\item the flat and rapidly rotating old thin disk,
\item the extended thick disk, including its metal weak tail,
\item a flattened and slowly rotating inner halo,
\item a spheroidal non-rotating outer halo,
\item satellite debris and kinematics substructures.
\end{enumerate}

\begin{figure}
 \plotone{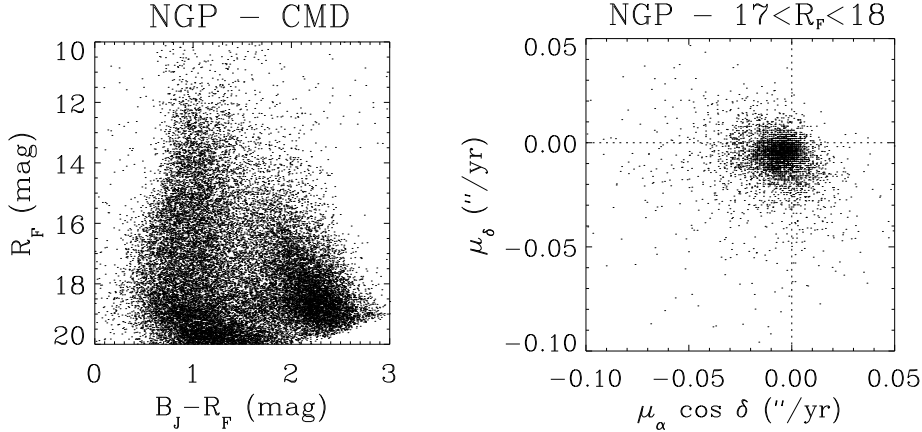}
 \caption{Color magnitude diagram ({\it left panel}) and vector point diagram ({\it right panel})
 for stars with $17\le R_F \le 18$. Both plots are based on data from the POSS-II field no.\ 442.}
\end{figure}

Actually, the physical properties of these components are not
completely established and various problems are still
controversial. For instance: $(a)$ the density scale factor,
rotation and metallicity distribution of the thick disk; $(b)$ the
nature of the metal weak thick disk (MWTD) and its relation with
the standard thick disk (satellite debris or initial phase of a
dissipative formation?); $(c)$ the halo velocity ellipsoid, the
spatial distribution and axial ratio of the halo as a function of
the distance, $(d)$ the search of halo streams; $(f)$ the
determination of the luminosity and mass function of the faintest
Pop.II stars (white dwarfs, late M dwarfs and subdwarfs).

\begin{figure}[t]
 \plotone{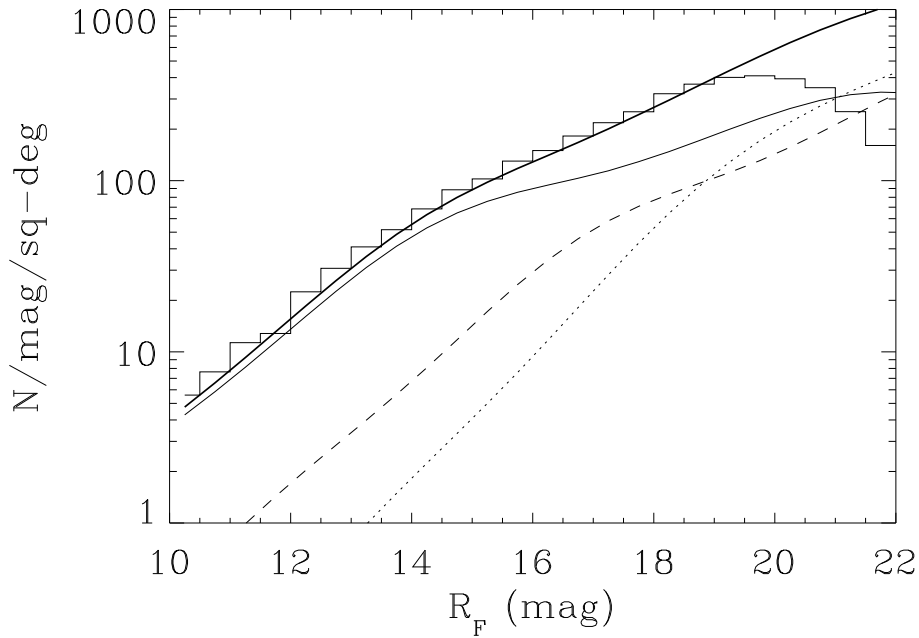}
 \caption{Starcounts derived from plate XP444 ({\it histogram})
 and compared with the distribution predicted by the Galaxy model ({\it thick solid line}),
 which includes thin disk ({\it solid  line}), thick disk ({\it dashed
  line}) and halo ({\it dotted line}) components. }
\end{figure}

Due to the lack of trigonometric parallaxes for our and similar
surveys, accurate distances and tangential velocities cannot be
directly determined. In such a case it is convenient to analyze
these large data-sets by comparing the observations against {\it
ad hoc} Galaxy models which describe both the density and
kinematics, as for instance the Besan\c{c}on model (Robin \& Oblak
1987), the IAS Galaxy model (Bachall, Casertano \& Ratnatunga
1987),  or the models developed by  Mendez \& van Altena (1996)
and Chen (1997).
 To this regard, in Figure 3 starcounts are
compared against the Mendez's Galaxy model which has been extended
to the photographic B$_J$ and R$_F$ magnitudes by Spagna (2001).

Alternatively, using tracers with known brightness, it is possible
to derive distance and space velocity by means of photometry,
proper motions and spectroscopic radial velocities.   This
approach has been adopted to investigate the kinematics of the
outer halo traced by a set of 31 RR Lyr{\ae} and 65 BHB giants, as
will be discussed in the following section.  These objects are
distributed between $2\la Z\la 12$ kpc (V=12-16.5 mag) and
accurate photometric parallaxes ($ \sigma_d/d\la 10$\%) have been
computed by means of the M$_V$ vs.\ B--V relation for BHB giants
derived by Preston et al.\ (1991), while for RR Lyr{\ae} stars we
adopted the $M_V$ as a function of metallicity or of Fourier
components, as follows:
\begin{eqnarray}
 M_V &=& 0.23\, \hbox{\rm [Fe/H]} +0.92 \\
 M_V &=&  -1.876 \log P -1.158 A_1 + 0.821 A_3 + 0.448
\end{eqnarray}
where Eq.\ 1 is from Cacciari (2002) and Chaboyer (1999), while
Eq.\ 2 is based on the relation from Kov\'{a}cs \& Walker (2001)
where  $P$ is the period (in days), $A_1$ and $A_3$ are the
Fourier amplitudes (mag) of the fundamental and second harmonic
components, respectively. The zero point has been calibrated by
Kinman (2002) with respect to the absolute magnitude
($M_V=0.61_{-0.11}^{+0.10}$) of RR Lyr derived from the HST/FGS
parallax measured by Benedict et al.\ (2002).
 Finally,  the extinction E(B--V) has been estimated from the maps
of Schlegel et al.\
 (1998).

\section{Halo rotation}
 A retrograde rotation of the outer halo has been suggested
by Majewski (1992) who measured a mean velocity $\langle V\rangle
= -275$ km s$^{-1}$, which corresponds to a galactocentric
retrograde velocity $v_{\rm rot}\simeq -55$ km s$^{-1}$ adopting
$V_{\rm LSR}=220$ km s$^{-1}$,
 from the analysis
of a pure sample of halo subdwarfs at $Z>5.5$ kpc towards the NGP.
 As shown in Table 2, this parameter
is still controversial, in fact Carney (1999), after correcting
the kinematics bias of his kinematically-selected subdwarf sample,
found a net prograde rotation of about $\langle V\rangle = -196$
km s$^{-1}$. On the contrary, Chiba \& Beers (2000) measured a
prograde rotating inner halo, $v_{\rm rot}\simeq 20$-60 km
s$^{-1}$, up to about 1 kpc, with a decreasing vertical gradient
of $dV/d|Z|=-52 \pm 6$ km s$^{-1}$ kpc$^{-1}$, while they did not
detect any significant rotation above $Z\sim 1.2$ kpc for very low
abundance stars ($-2.4\le$ [Fe/H] $\le -1.9$), where contamination
of thick disk stars should be negligible.  In addition, their halo
sample at larger distances (212 stars with $4<Z_{\rm Max}< 20$ kpc
and [Fe/H]$\le -1.5$) still does not support any significant
rotation: $v_{\rm rot}\simeq 0 \pm 8$ km s$^{-1}$.

\begin{table}
  \centering
  \caption{Recent measurements of the halo rotation.}
\begin{tabular}{lccr}
  &  &  &  \\
 \hline
   &  &  &  \\
  Tracers & N.ro Objects & $\langle V\rangle$ (km/s)& Reference\\
   \vspace{0.1cm} \\
     \tableline
     \multicolumn{4}{c}{ \sc Inner Halo} \\
 RR Lyr{\ae}  & 162 ($|Z|<2 $ kpc) & $-210\pm 12$* & Layden et al.\ (1996) \\
 RR Lyr{\ae}  & 84 ($|Z|<2 $ kpc) & $-219\pm 10$*  & Martin \& Morrison (1998) \\
 RR Lyr{\ae} & 147 ($|Z|<2$ kpc)    & $-217\pm 13$*  &  Gould \& Popowski (1998) \\
  Subdwarfs  &  ($|Z|<1$ kpc)& $-(160\div 200)$** & Chiba \& Beers (2000) \\ 
 \tableline
  \multicolumn{4}{c}{ \sc Outer Halo}\\
 Subdwarfs &  21 ($Z>4$ kpc)    & $-275\pm 16$  & Majewski (1992) \\
 Subdwarfs & 30 ($Z_{\rm Max}>5$ kpc)    & $-196\pm 13$   & Carney (1999)\\
 Subdwarfs &  212 ($Z_{\rm Max}>4$ kpc)    & $-220\pm 8$   & Chiba \& Beers (2000)\\ 
 RR \& BHB & 53 ($2<Z<12$ kpc) &     $-285\pm 17$*$^,$***    &  This
 survey  \\
 \smallskip\\
 \tableline
  \multicolumn{4}{l}{(*) Heliocentric velocities. (Wrt. the LSR in the other cases.)}\\
  \multicolumn{4}{l}{(**) As a function of distance with a gradient $dV/d|Z|=-52 \pm 6$
   km s$^{-1}$ kpc$^{-1}$ }\\
   \multicolumn{4}{l}{\hspace{0.5cm} ($-2.4<$[Fe/H]$<-1.9$).}\\
   \multicolumn{4}{l}{(***) Preliminary results based on
   18 RR Lyr{\ae} stars and 35 BHB giants).}
\end{tabular}
  \label{rotation}
\end{table}

However, recently Gilmore et al.\ (2002), who carried out a
spectroscopic survey at intermediate galactic latitudes  of about
2000 F/G stars, revealed a significant excess of {\it retrograde}
halo stars in their faintest magnitude bin ($18<V<19.5$)
corresponding to a vertical distance $|Z|\approx 5$ kpc.

The fact that the velocities of halo stars do not match an exact
gaussian distribution is well known (see eg.\ Martin \& Morrison
1998).  How much this depends on the properties of the whole
population or on the effects of kinematic substructures, such as
satellite debris of ancient accretion events (eg.\ Helmi et al.\
1999), remains to be established.

As shown in Table 2, a preliminary analysis of our sample of RR
Lyr{\ae} and BHB giants (Kinman et al.\ 2002) seems to support a
retrograde rotation of the outer halo.  In fact, we measured a
heliocentric velocity $\langle V\rangle = -285 \pm 17$ km
s$^{-1}$, which corresponds to $v_{\rm rot}\simeq -60$  km
s$^{-1}$ adopting a solar motion with respect to the LSR,
$V_\odot=+5.25\pm 0.62$ km s$^{-1}$,  from Dehnen \& Binney (1998)
and assuming $V_{\rm LSR}=220$  km s$^{-1}$.   This result is
confirmed also by the separate analysis of the RR Lyr{\ae} and BHB
giants, which both provide a retrograde rotation.

How much this value may be affected by a velocity bias can be
estimated by the level of systematic errors on proper motions
which in practice give the main contribution to the $U$ and $V$
galactic components along this line of sight towards $b\approx
90^\circ$.  As reported in Sect.\ 2, we found systematic errors
$\Delta\mu < 1$ mas/yr, from which velocity bias up to 20-30 km
s$^{-1}$ can be expected for such stars located at $Z\approx 5$-6
kpc, on average. Clearly, this is a critical point which need
further analysis.

\begin{figure}[tb]
 \plotfiddle{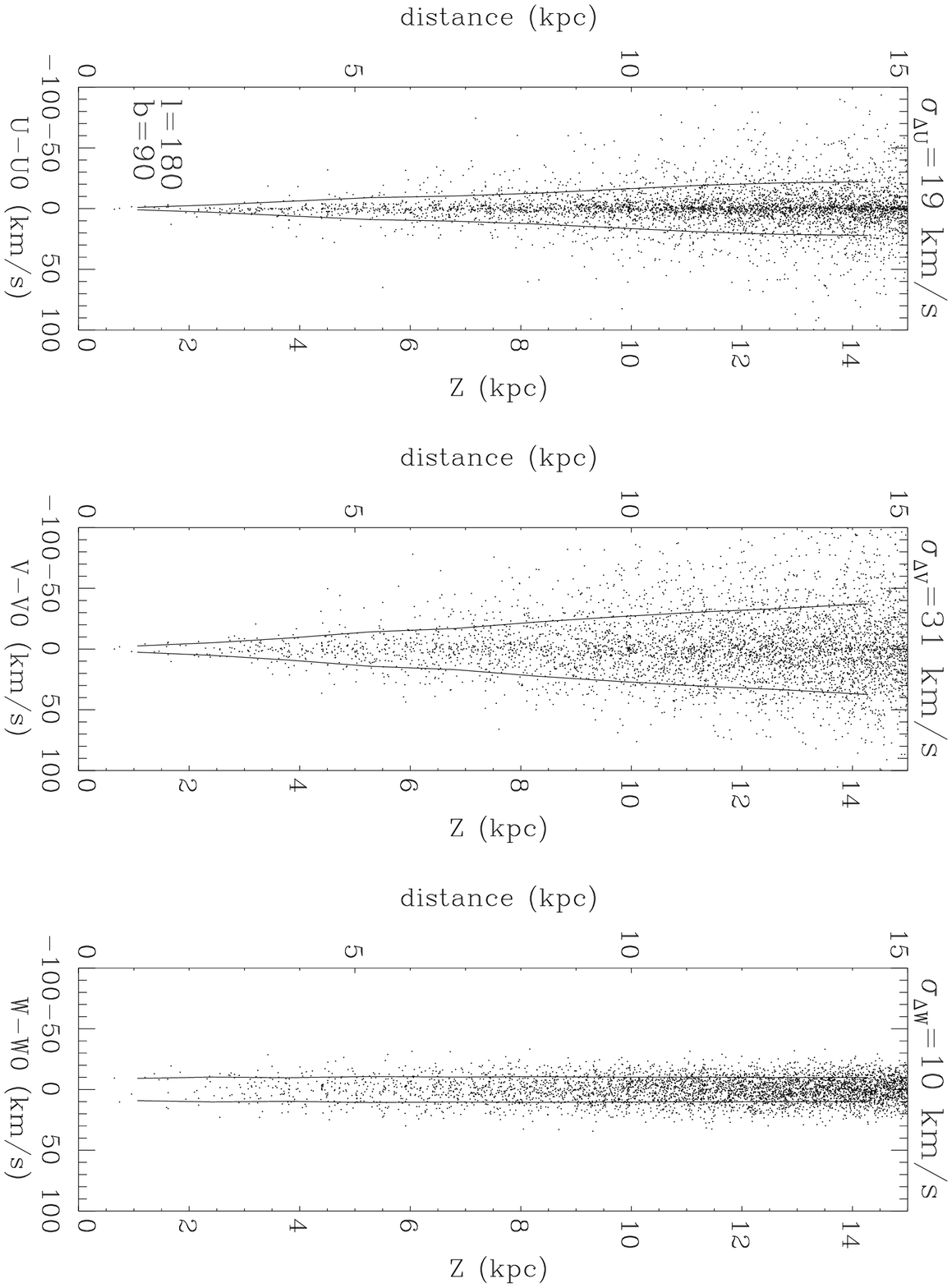}{8.5cm}{90}{60}{50}{220}{-20}
 \plotfiddle{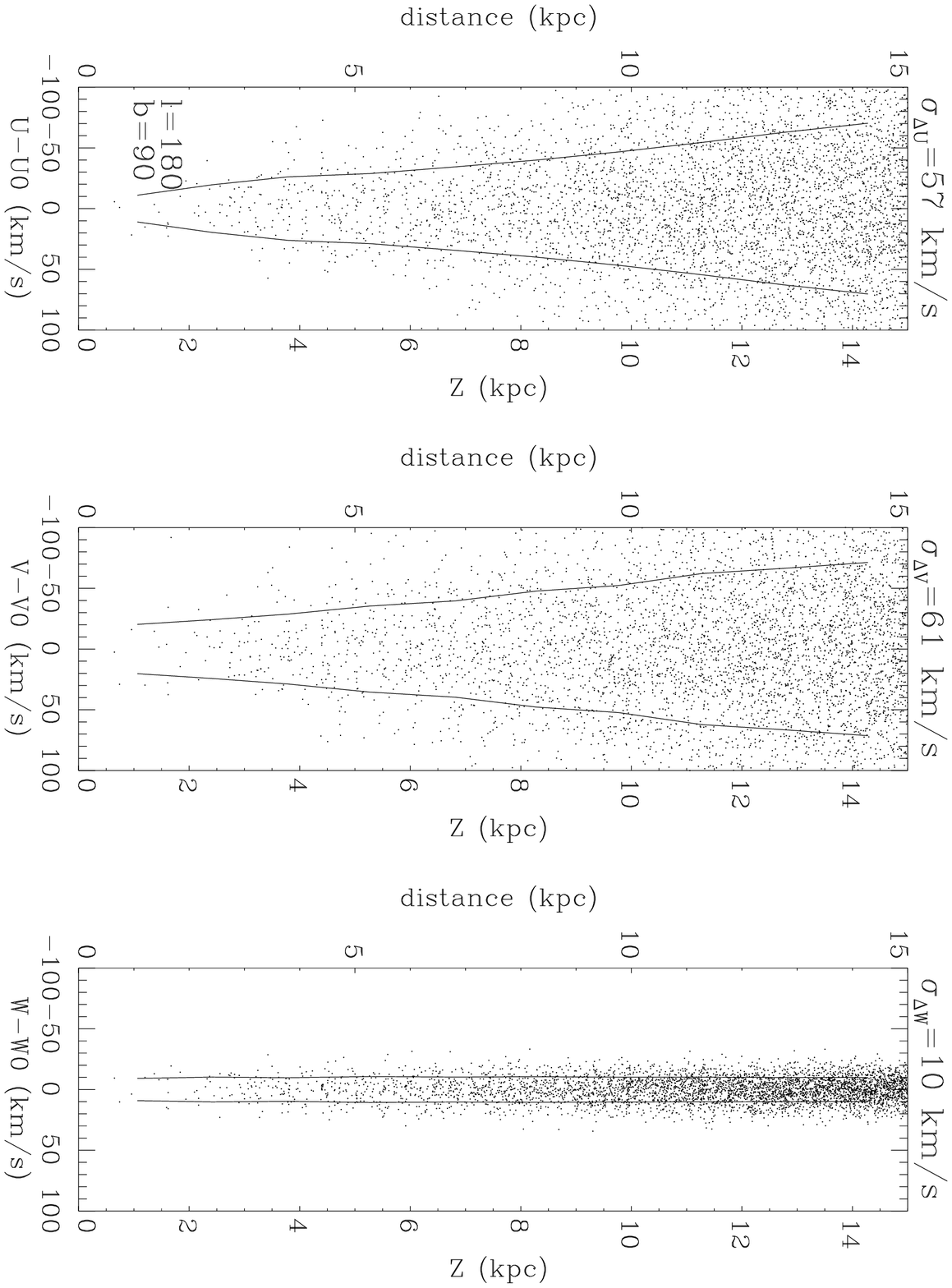}{8.5cm}{90}{60}{50}{220}{-20}

  \caption{Kinematics simulation towards NGP. Velocity residuals
  for Case A ({\it top panels}) and Case B ({\it bottom panels}).
   Solid lines show the precision level ($\pm 1 \sigma$) as function of the distance
   up to Z=15 kpc. Above each plot, the rms of the residuals of the whole sample
   is reported.}
\end{figure}

\section{Kinematics simulation}
A Montecarlo simulation has been developed in order to compare the
GAIA capability to recover the halo kinematics with respect to the
velocity precision that can be attained by the current and future
ground based surveys. For simplicity we considered an uniform
spatial distribution with Pop.II-like kinematics, $(\sigma_U,
\sigma_V,\sigma_W) =(150, 100, 100)$ km~s$^{-1}$ and an observer
rotating with a velocity of 220 km~s$^{-1}$ at a distance of 8 kpc
from the galactic center. Implicitly we assumed to measure bright
tracers, such as RR Lyr{\ae}, BHB and red giants, having apparent
magnitude $V\sim 16$ mag at a distance of about $d\sim 10$ kpc.
The following cases were tested:
\begin{itemize}
 \item Case A (GAIA), with $\sigma_\pi=10$ $\mu$as, $\sigma_\mu=10$ $\mu$as~yr$^{-1}$
  (per component) and $\sigma_{\rm Vr}=10$ km~s$^{-1}$,
 from astrometric and spectroscopic observations;
 \item Case B (ground-based surveys), with $\sigma_{m-M}=0.2$ mag, $\sigma_\mu=1$ mas~yr$^{-1}$
  (per component) and $\sigma_{\rm Vr}=10$ km s$^{-1}$, from
  astrometric, photometric and spectroscopic observations.
\end{itemize}
Note that the two cases differ essentially in a factor 100 on the
proper motion accuracy. In fact, at $\sim 10$ kpc the distance
accuracy is the same for both cases.  This points out the fact
that, although radial velocities and distance moduli derived from
ground based spectro-photometric surveys can attain similar
accuracy, GAIA astrometry will uniquely provide accurate and
reliable tangential velocities up to large distances.

Figure 4 shows the results of Montecarlo simulations with 5000
stars towards $b=90^\circ$. Top panels present the velocity errors
(i.e.\ the differences between the observed velocity and the true
value) for Case A, and those at the bottom for Case B. As
expected, errors increase linearly with distance, and for GAIA the
rms of $\Delta U$ and $\Delta V$ residuals vary from $\sim 10$
km~s$^{-1}$ at 5 kpc to 20-30 km~s$^{-1}$ at 10 kpc. For Case B
the errors on U and V are about a factor 2-3 larger, while
$\sigma_{\Delta W}= \sigma_{\rm Vr} \equiv 10$ km~s$^{-1}$ in both
the cases.   Note that in Case A the error on the tangential
velocity is in practice dominated by the distance uncertainty,
$\sigma_\pi$, while both the errors on proper motions and
photometric parallaxes contribute to the velocity errors of Case
B. Similar results are derived in other directions.

These velocity errors should be compared with the typical motion
of the halo stars (100-200 km~s$^{-1}$) for which GAIA will
provide individual 3D spatial velocities with a significant
signal-to-noise, $\sigma_v/v$, up to $d \approx 10$-15 kpc, a
distance where ground based surveys are not able to measure
reliable tangential velocities.
 GAIA will measure direct distances
and velocities for large samples of halo tracers, selected {\it in
situ} without kinematics nor metallicity bias, from which it will
possible to determine accurately the halo velocity ellipsoid and
its orientation.  In particular, the GAIA $\mu$-arcsec level
accuracy will be essential in order to take advantage of the
$\sqrt{N}$ statistical factor and avoid the risk of velocity
biases due to the presence of systematics errors affecting proper
motions and parallaxes as discussed in Sect.\ 4.

Finally, accuracy of the order of 15-20 km~s$^{-1}$, such as that
attained by GAIA for $d \la 10$ kpc, is the value requested to
resolve kinematics substructures in the halo, as the satellite
debris predicted by the hierarchical scenarios of galaxy
formation.  To this regards, Helmi (2002) estimated that GAIA will
be able to recover 2/3 of the accretion events with a velocity
accuracy better than 20 km~s$^{-1}$ and  $\sigma_d/d<$ 20\% down
to $V\simeq 18.5$, or 1/2 of the events down to $V\simeq 15$ mag.

\section{Summary}
Large field proper motion surveys ($\sigma_\mu\approx$ 1-10
mas/yr) based on photographic surveys digitized with fast
measuring machines (APS, MAMA, PDS, PMM, SuperCOSMOS) are still
useful tools for the study of the structure and kinematics of the
galactic stellar populations, especially when combined with radial
velocities and chemical abundances from spectro-photometric
observations.
 In particular, we have shown how bright halo tracers, such as BHB
 giants and RR Lyr{\ae} stars with $V\la 16$-17, can be used to
 investigate {\it in situ} the halo
 kinematics up to $z\approx 5$-6 kpc via photometric parallaxes
 if proper motions with
 $\sigma_\mu \sim 1$-2 mas/yr accuracy and radial velocities with
precision of about $\sigma_{\rm Vr}\sim 10$-20 km~s$^{-1}$ are
available.  For statistical analysis of large samples, systematic
errors on the proper motions (and $M_V$ too) are critical and can
result in biased mean velocities and dispersions.
 These problems will be dramatically reduced by the astrometric
 parameters and spectroscopic $V_r$  provided by GAIA, which will permit
 to determine
 $(a)$ direct distances from parallaxes, $(b)$ radial and tangential velocities with an
accuracy of a few tens of km~s$^{-1}$ to $d\sim 10$-15 kpc for
$(c)$ unbiased and larger sets of tracers identified by means of
its spectro-photometric and astrometric data.

\acknowledgments
 The GSC II is a joint project of the Space Telescope
Science Institute and the Osservatorio Astronomico di Torino.
Space Telescope Science Institute is operated by AURA for NASA
under contract NAS5-26555. Partial financial support to this
research comes from the Italian CNAA and the Italian Ministry of
Research (MIUR) through the COFIN-2001 program.

\end{document}